\documentclass{aa}

\usepackage{amsmath}
\usepackage{txfonts}
\usepackage{graphicx}
\usepackage{natbib}
\bibpunct{(}{)}{;}{a}{}{,} 

\newsavebox{\obox}

\title{A hydrodynamic study of the circumstellar envelope of \object{$\alpha$~Scorpii}\thanks{Based
       on observations collected at the European Organisation for Astronomical Research in the
       Southern Hemisphere, Chile, under program ID 076.D-0690(A), and on observations made with the
       NASA/ESA Hubble Space Telescope (program \# 5952), obtained at the Space Telescope Science
       Institute, which is operated by the Association of Universities for Research in Astronomy,
       Inc., under NASA contract NAS 5-26555.}}
\author{K.~Braun \and R.~Baade \and D.~Reimers \and H.-J.~Hagen}
\institute{Hamburger Sternwarte, Universit\"at Hamburg, Gojenbergsweg 112, 21029 Hamburg,
           Germany\newline\email{rbaade@hs.uni-hamburg.de}}
\date{Received 23 Mai 2012 / Accepted 24 August 2012}

\abstract
{Both the absolute mass-loss rates and the mechanisms that drive the mass loss of late-type
 supergiants are still not well known. Binaries such as $\alpha$~Sco provide the most detailed
 empirical information about the winds of these stars.}
{The goal was to improve the binary technique for the determination of the mass-loss rate of
 \object{$\alpha$~Sco~A} by including a realistic density distribution and velocity field from
 hydrodynamic and plasma simulations.}
{We performed 3D hydrodynamic simulations of the circumstellar envelope of $\alpha$~Sco in
 combination with plasma simulations accounting for the heating, ionization, and excitation of the
 wind by the radiation of \object{$\alpha$~Sco~B}. These simulations served as the basis for an
 examination of circumstellar absorption lines in the spectrum of $\alpha$~Sco~B as well as of
 emission lines from the \object{Antares nebula}.}
{The present model of the extended envelope of $\alpha$~Sco reproduces some of the structures that
 were observed in the circumstellar absorption lines in the spectrum of $\alpha$~Sco~B. Our
 theoretical density and velocity distributions of the outflow deviate considerably from a
 spherically expanding model, which was used in previous studies. This results in a higher mass-loss
 rate of $(2\pm0.5)\times10^{-6}\ M_{\odot}\,\mathrm{yr}^{-1}$. The hot \ion{H}{ii} region around
 the secondary star induces an additional acceleration of the wind at large distances from the
 primary, which is seen in absorption lines of \ion{Ti}{ii} and \ion{Cr}{ii} at $-30\
 \mathrm{km\,s}^{-1}$.}
{}


\begin{document}

\maketitle

\section{Introduction} \label{introd}
In the late stage of the stellar evolution stars suffer a substantial mass loss by a continuous
wind. This phenomenon plays a key role for the final development of individual stars and for the
enrichment of the interstellar medium. However, the physical mechanisms that govern the outflow of
late-type giants and supergiants are not established. Possible driving mechanisms include Alfv\'{e}n
waves, acoustic waves, shock waves, and pulsations \citep[see the reviews of][]{Lafon_Berruyer_1991,
Willson_2000}. As it is hard to distinguish between different driving mechanisms from a theoretical
point of view, there is still a need for empirical studies to provide constraints.

The most accurate mass-loss rates are obtained from observations of eclipsing binaries of
$\zeta$~Aur or VV~Cep type. These systems consist of a late-type supergiant and a main-sequence
companion that can be used as a ``natural satellite'' to probe the circumstellar envelope. For the
basic concepts of the binary technique see e.\,g.\ \citet{Reimers_1987},
\citet{Hack_Stickland_1987}, or \citet{Baade_1998}. Observations with high spectral resolution of
$\zeta$~Aur \citep{Baade_etal_1996} and $\alpha$~Sco \citep{Baade_Reimers_2007, Reimers_etal_2008}
have shown that the circumstellar envelopes of these stars cannot be described by a continuous,
spherically expanding model. One of the aims of this study was therefore to investigate the
influence of the binary companion on the density and velocity distribution of the supergiant wind.

The $\alpha$~Sco system (M1\,Ib + B2.5\,V) is of particular importance in this context, because its
angular separation amounts to nearly 3\arcsec, which allows the observation of the extended envelope
with spatial resolution. \citet{Reimers_etal_2008} presented the most recent extensive observations,
which were carried out with the Ultraviolet and Visual Echelle Spectrograph (UVES) on the Kueyen
telescope of the Very Large Telescope array (VLT) and provide spatially resolved emission spectra of
the \object{Antares} nebula. They showed that the mass-loss rate of $\alpha$~Sco~A can be determined
by measuring the spatial extent of the H$\alpha$ emission. They determined a mass-loss rate of
$\dot{M}=(1.05\pm0.3)\times10^{-6}\ M_\odot\,\mathrm{yr}^{-1}$ as a best match between the
measured H$\alpha$ distribution and plasma simulations, which were based on a spherically expanding
model of the circumstellar envelope and led to a nearly spherical \ion{H}{ii} region around the
secondary that is rotationally symmetric with respect to the line connecting the two stars. They
reported an apparent increase of the mass-loss rate towards larger distances from the supergiant.
Moreover, they noted that the observed [\ion{Fe}{ii}] emission cannot be explained in the framework
of their simplified model.

Mid-infrared observations at the Keck~\textsc{ii} telescope of the dust in the circumstellar
envelope of $\alpha$~Sco indicate a non-uniform dust distribution that appears to be related to
discrete ejections of mass from $\alpha$~Sco~A \citep{Marsh_etal_2001}. This is consistent with
interferometric observations that were carried out at the William Herschel Telescope by
\citet{Tuthill_etal_1997} and revealed asymmetric structures (``hotspots'') at the surface of the
supergiant.

\citet{Baade_Reimers_2007} presented spectra of $\alpha$~Sco~B, which were obtained with the Goddard
High-Resolution Spectrograph (GHRS) on the Hubble Space Telescope (HST), including a number of
circumstellar absorption lines. These absorption lines exhibit multi-component structures that are
not compatible with a continuous wind, and the derived column densities indicated a complex
differential depletion as a result of dust formation. They showed that these structures can be
explained in principle either by a model consisting of concentric shells with different densities,
consistent with a highly variable mass-loss rate, or by discrete clumps in the line of sight due to
unknown ejection processes.

The goal of this work was to construct a more realistic model of the $\alpha$~Sco system, including
hydrodynamic and plasma effects, in order to gain quantitative insight into the dynamics of the
circumstellar envelope. After a description of our hydrodynamic and plasma simulations of the
$\alpha$~Sco system and its \ion{H}{ii} region in Sect.~\ref{modcal}, we present a comparison to
observations in Sect.~\ref{result}.

\section{Model calculations} \label{modcal}
Our model includes the hydrodynamic effects of the hot \ion{H}{ii} region around the secondary star,
$\alpha$~Sco~B, on the density and velocity distribution of the circumstallar envelope of the binary
star. The temperature and ionization inside the \ion{H}{ii} region are calculated by considering
both microphysical and radiative processes (see Sect.~\ref{plasim}), and the result is used as input
into the hydro code (Sect.~\ref{hydsim}).

\subsection{Stellar and system parameters}
$\alpha$~Sco is a visual binary system consisting of an M1\,Ib-type red supergiant and a
B2.5\,V-type dwarf, and its distance is $d=185$~pc according to the Hipparcos catalogue.
\citet{Hopmann_1958} determined the inclination of the orbit and found $i\sim90\degr$. The angular
separation is 2\farcs73 and the B star is $\sim224$~AU behind the supergiant, which corresponds to a
position angle $\delta=23\pm5\degr$ \citep{Reimers_etal_2008}.

According to \citet{Kudritzki_Reimers_1978} the effective temperature of the B star is
$T_\mathrm{eff, B}=18\,500\pm1500$~K, and its surface gravity is $\log{g_\mathrm{B}}=3.9\pm0.2$ (in
cgs units). These values result from profile fits of Balmer and \ion{He}{i} lines, and the mass
$M_\mathrm{B}$ of the B star was determined by a comparison of these values with theoretical
evolutionary tracks, yielding $M_\mathrm{B}=7.2\pm0.5\ M_\odot$ and $R_\mathrm{B}=5.2\pm1.3\
R_\odot$. Applying more recent evolutionary models \citep{Bressan_etal_1993} gives similar results,
i.\,e.\ $M_\mathrm{B}=6.7\pm0.7\ M_\odot$ and $R_\mathrm{B}=4.8\pm1.1\ R_\odot$.
\citet{Brott_etal_2011} presented calculations including rotational effects. Their evolutionary
sequence for $7\ M_\odot$ and $\varv_\mathrm{rot}=284\ \mathrm{km\,s}^{-1}$ matches the surface
gravity and effective temperature derived for $\alpha$~Sco~B by \citet{Kudritzki_Reimers_1978}, who
observed a projected rotational velocity of $\varv_\mathrm{rot}\sin{i}=250\ \mathrm{km\,s}^{-1}$.
Based on these results and the analysis of \citet{Hjellming_Newell_1983}, who gave a constraint on
the Lyman continuum luminosity of the B star, we adopted $T_\mathrm{eff, B}=18\,200$~K.

For the mass of $\alpha$~Sco~A we adopted the value $M_\mathrm{A}\sim18\ M_\odot$ found by
\citet{Kudritzki_Reimers_1978}. Table~\ref{syspar} lists the parameters used in this work.

\begin{table}
 \caption{System parameters of $\alpha$~Sco used in this work.}
 \label{syspar}
 \centering
 \begin{tabular}{l l l}
  \hline\hline
  Parameter & Value & Source \\
  \hline
  $T_{\mathrm{eff, B}}$       & 18\,200~K     & This work                      \\
  $\log{g_\mathrm{B}}$        & 3.9 (cgs)     & \citet{Kudritzki_Reimers_1978} \\
  $M_\mathrm{B}$              & 6.7~$M_\odot$ & This work                      \\
  $M_\mathrm{A}$              & 18~$M_\odot$  & \citet{Kudritzki_Reimers_1978} \\
  $d$                         & 185~pc        & Hipparcos catalogue            \\
  $\delta$                    & $23\pm5\degr$ & \citet{Reimers_etal_2008}      \\
  Separation (in 2006)        & 2\farcs 73    & \citet{Reimers_etal_2008}      \\
  $\sin{i}$                   & 1             & \citet{Hopmann_1958}           \\
  $e$                         & 0             & \citet{Hopmann_1958}           \\
  \hline
 \end{tabular}
 \tablefoot{$\delta$ is the position angle of the B star with respect to the primary (see text). $i$
            and $e$ are the inclination and eccentricity of the orbit, respectively.}
\end{table}

\subsection{Hydrodynamic simulations} \label{hydsim}
For the hydrodynamic simulations we used the hydro version (without magnetic fields) of the
AMRCART code, which is part of the A-MAZE package \citep{Walder_Folini_2000} and comprises the
adaptive mesh refinement algorithm developed by \citet{Berger_Colella_1989}. AMRCART solves
the Euler equations in 3D and uses the equation of state of an ideal gas,
\begin{equation}
 e=\frac{p}{(\gamma-1)\rho},
\end{equation}
where $e$ is the specific internal energy, $p$ the pressure,
$\rho$ the mass density, and $\gamma=5/3$. We used the finite-volume
integrator that is based on a modified Lax-Friedrichs approach \citep{Barmin_etal_1996}, which is
implemented in the AMRCART code. This approach uses a predictor-corrector procedure to
obtain a scheme with a second-order accuracy. We did not include the gravitational potential of the
B star, because we were interested in the large-scale effects that are caused by the heating that is
due to its radiation, and the resolution of the computational grid is not sufficient for a
simultaneous calculation of possible accretion effects. The heating by the radiation of the B star
is included as a source term in the Euler equations (see Sect.~\ref{plasim}).

\subsubsection{Boundary and initial conditions}
We assume that the wind is rapidly accelerated to its terminal velocity $\varv_\infty$ near the
supergiant. This assumption does not have a significant influence on the results, because the region
of interest, where the wind interacts with the \ion{H}{ii} region, is far from the supergiant, at a
distance of the order of $\sim300$~AU. Thus, we simply apply the terminal velocity at the boundary
cells, i.\,e.
\begin{equation} \label{veloci}
 \vec{\varv}(R_\mathrm{A})=\vec{\varv}_\infty+\vec{\varv}_\mathrm{A},
\end{equation}
where $R_\mathrm{A}$ is the radius and $\vec{\varv}_\mathrm{A}$ the orbital velocity of the primary
star, and $\vec{\varv}_\infty$ points radially outward from the primary.

At the outer boundaries of the computational domain ``free flow'' is assumed, i. e., density and
velocity in boundary grid cells equal the values of the cells at the edge of the domain. The initial
velocity field is given by Eq.~\ref{veloci} at all radii $r\geq R_\mathrm{A}$, and the initial
density distribution is spherically symmetric with respect to the primary, i.\,e.
\begin{equation}
 \rho(r)=\frac{\dot{M}}{4\pi r^2\varv_\infty}.
\end{equation}

\subsubsection{The computational grid}
As the system is symmetric with respect to the plane of the orbit, we place the center of mass near
the upper boundary of the computational domain, which reduces the computing time by reducing the
total number of grid cells. The plane of the orbit is parallel to the $xy$ plane, and the center of
mass of the system is placed in the center of that plane. For most of the simulations presented in
this work, two levels of refinement are used with a refinement ratio of 2 between the respective
grid spacings. The base grid comprises $160\times160\times44$ grid cells corresponding to a size
of $a_x\times a_y\times a_z\sim12\,433\times12\,433\times3419$~AU, where $a_x$, $a_y$, and $a_z$ are
the edge lengths of the domain in $x$, $y$, and $z$ direction, respectively.

For the refinement of the grid, the local truncation error
$\mathit{TE}(\vec{r},t)$ of the density is estimated for each cell of the base grid with
a method that is based on Richardson extrapolation \citep[see][]{Berger_Colella_1989}. If the
truncation error of a grid cell exceeds a given tolerance limit $\varepsilon$, the grid
cell is flagged. In the next step, cuboid-shaped subgrids are generated that include all regions
containing flagged cells. We used an error tolerance of $\varepsilon=4\times10^{-4}$ and
enhanced the resolution by a factor of two in the subgrids in our simulations. The error tolerance
was applied to the relative truncation error, i.\,e.\
$\mathit{TE}(\vec{r},t)/\rho(\vec{r},t)\leq\varepsilon$. The contours of the subgrids are
indicated as gray rectangles in Fig.~\ref{denvel}.

For the calculation of the heating by the radiation of the B star (see next section),
spherical coordinates are used, which are then converted to the cartesian grid of the hydro code by
linear interpolation. In Appendix~\ref{clodis} we describe the generation of the grid that provides
the temperature around the B star as a function of position in 3D space.

\subsection{Plasma simulations} \label{plasim}
The temperature and ionization inside the \ion{H}{ii} region around the B star were calculated with
version 08.00 of the plasma code Cloudy, last described by \citet{Ferland_etal_1998}. Cloudy uses a
spherical geometry in 1D. In order to obtain a 3D temperature distribution of the $\alpha$~Sco
system we calculated different models, each one corresponding to a different direction, starting
from the position of the source of ionizing radiation, i.\,e.\ the B star (see
Appendix~\ref{clodis} for details). This approach is justified as Cloudy uses the escape probability
formalism for the calculation of radiative transfer effects, so that different points inside the
circumstellar envelope do not interact with each other. For the radiation of the B star we used a
synthetic spectrum given by an interpolation on the TLUSTY grid of \citet{Lanz_Hubeny_2007}.

The resulting temperature distribution is then passed to the hydro code as a source term
affecting the total energy. The radiative heating is assumed to take place at a much smaller time
scale than the hydrodynamic processes. This procedure is repeated at regular intervals during the
calculation, typically every other time step as measured on the base grid. The information about the
state of ionization can be used later in the analysis of spectral lines.

For the treatment of \ion{Fe}{ii}, Cloudy provides a large model atom developed by
\citet{Verner_etal_1999}. It includes the 371 lowest levels of the Fe$^{+}$ ion, i.\,e., up to
93\,487.65~cm$^{-1}$.

\section{Results} \label{result}

\subsection{Density and velocity distribution} \label{hydeve}
\begin{figure}
 \resizebox{\hsize}{!}{\includegraphics{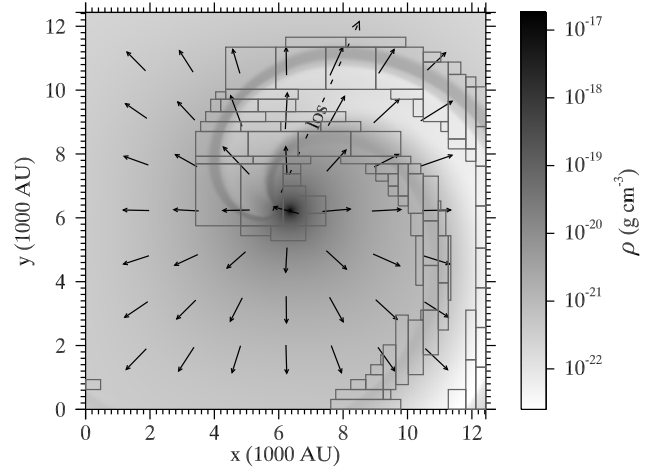}}
 \caption{Quasi-stationary density and velocity distribution resulting from a simulation
          with $\dot{M}=2\times10^{-6}\ M_\odot\,\mathrm{yr}^{-1}$ and $\varv_{\infty}=20\
          \mathrm{km\,s}^{-1}$ after $\sim1$~orbit or $\sim2560$~yrs.
          The dotted arrow indicates the line of sight to the B star, whose position is marked by
          the asterisk. The gray rectangles indicate the contours of the high-resolution
          subgrids (see text).}
 \label{denvel}
\end{figure}

\begin{figure}
 \resizebox{\hsize}{!}{\includegraphics{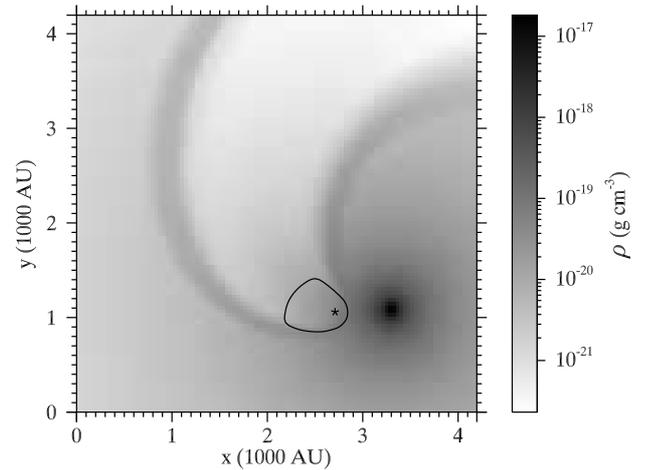}}
 \caption{Enlarged central region of Fig.~\ref{denvel}. The contour indicates 90\% ionization of
          hydrogen.}
 \label{denzoo}
\end{figure}

The presence of the hot \ion{H}{ii} region inside the wind of the supergiant causes deviations from
spherical symmetry. Figure~\ref{denvel} shows a cut through the center of mass of the system
parallel to the plane of the orbit, including the density distribution as well as appropriately
scaled velocity vectors for a simulation with $\dot{M}=2\times10^{-6}\
M_\odot\,\mathrm{yr}^{-1}$ and $\varv_{\infty}=20\ \mathrm{km\,s}^{-1}$ after
$\sim1$~orbit or $\sim2560$~yrs. At the boundaries of the \ion{H}{ii} region
around the B star that face the supergiant the density increases, and it decreases inside the wake
of the \ion{H}{ii} region. The curved shape of the structure results from the orbital motion. The
orientation of the velocity vectors is still approximately spherically symmetric, but the absolute
value increases by $\sim50$\% in the low-density regions. Figure~\ref{denzoo} shows the central
region of Fig.~\ref{denvel} including the \ion{H}{ii} region as a contour that indicates 90\%
ionization of hydrogen.

We calculated models with different values of $\dot{M}$ (0.2, 0.5, 1, 2, and $5\times10^{-6}\
M_\odot\,\mathrm{yr}^{-1}$) and $\varv_\infty$ (10, 20, and 40~km\,s$^{-1}$), including all
possible combinations of $\dot{M}$ and $\varv_\infty$ in these ranges.

\subsection{Comparison to observed circumstellar absorption lines}
Circumstellar absorption lines in the spectrum of $\alpha$~Sco~B have been observed with GHRS/HST
\citep{Baade_Reimers_2007} and with UVES/VLT \citep{Reimers_etal_2008}. In Sect.~\ref{synpro}, we
describe the calculation of synthetic line profiles based on the simulations presented above. These
synthetic profiles are compared to observations in Sect.~\ref{obspro}. For that comparison, we
calibrated the wavelength scale of the GHRS spectra by use of the UVES spectra (see
Sect.~\ref{calwav}). It should be noted that the resonance lines (0~eV) of abundant ions are
affected by interstellar contamination, as shown by \citet{Hagen_etal_1987}. Exceptions are those
elements which are heavily depleted in the IS medium.

\subsubsection{Synthetic line profiles} \label{synpro}
In order to derive the profiles of circumstellar lines we use the density and velocity information
from the hydrodynamic simulations to calculate a pure absorption line profile given by
\begin{equation}
 I_\nu=I_\mathrm{B}\exp{\left(-\int^{z_\mathrm{max}}_{z_\mathrm{B}}\chi_\nu\mathrm{d}z\right)},
\end{equation}
where $I_\nu$ is the emergent intensity, $I_\mathrm{B}$ the intensity of the B star continuum,
$\chi_\nu$ the opacity, and $z$ the coordinate along the line of sight towards the B star, which is
located at $z=z_\mathrm{B}$. $z_\mathrm{max}$ is chosen so large that the line profile does not
change if a higher value is used for this limit. After about one half of an orbit the
density structure becomes quasi-stationary and does not change its shape anymore. Therefore, the
results from the hydrodynamic simulations at any time step can be used for the calculation of the
line profile. We used the density information corresponding to the time after about one orbit,
i.\,e.\ at $t\sim P$, where $P\sim2560$~yrs is the orbital period of the
system.

\begin{figure}
 \resizebox{\hsize}{!}{\includegraphics{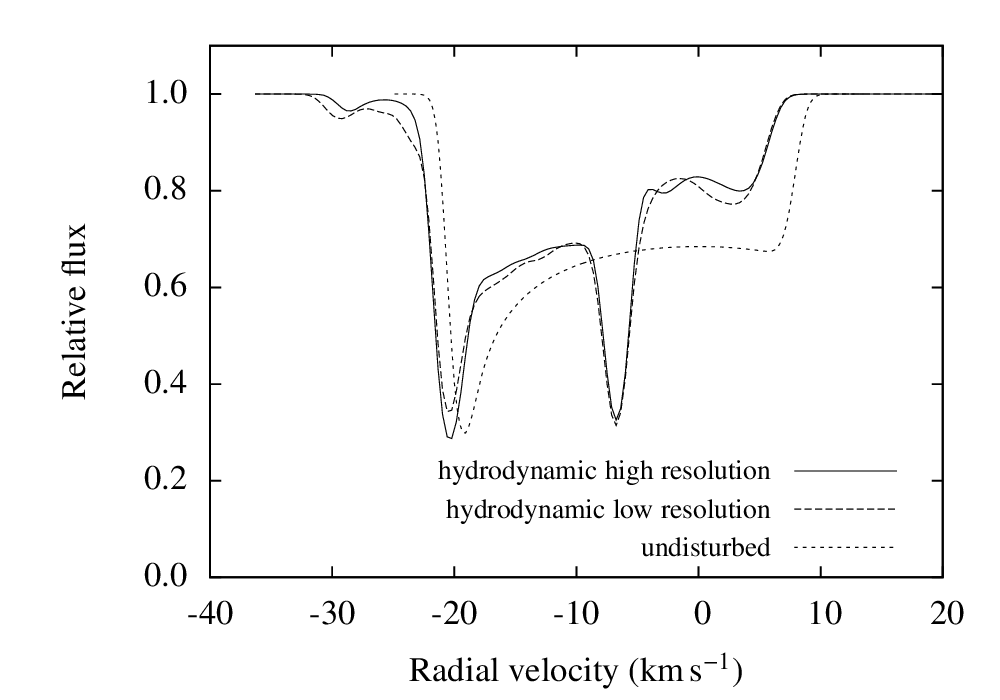}}
 \caption{Calculations of the \ion{Zn}{ii} UV mult.~1 2062.660~{\AA} absorption line based on a
          hydrodynamic simulation including ionization with $\dot{M}=10^{-6}\
          M_\odot\,\mathrm{yr}^{-1}$ and $\varv_{\infty}=20\ \mathrm{km\,s}^{-1}$, and on an
          undisturbed, spherically expanding wind. As a convergence test, the hydrodynamic
          results (\emph{low resolution}) are compared to a model with the same parameters but with
          a resolution that is increased by a factor of two (\emph{high resolution}). The difference
          in the component at $\sim-29\ \mathrm{km\,s}^{-1}$ is due to the reduced size
          of the computational domain in the high-resolution simulation (see text). The abscissa
          gives the radial velocity relative to the center of mass of the system.}
 \label{abspro}
\end{figure}

The resulting profiles deviate considerably from the profiles resulting from a continuous,
spherically symmetric model. As an example, Fig.~\ref{abspro} shows the simulated profile of the
\ion{Zn}{ii} UV mult.~1 2062.660~{\AA} absorption line, based on a simulation with $\dot{M}=10^{-6}\
M_\odot\,\mathrm{yr}^{-1}$ and $\varv_\infty=20\ \mathrm{km\,s}^{-1}$, along with a plot of the same
line for the case of an undisturbed, spherically expanding wind with the same parameters. Obviously,
the absorption component at $\sim-20\ \mathrm{km\,s}^{-1}$ is present in both lines and coincides
with the terminal wind velocity, but the absorption profile that is based on the hydrodynamic
simulation exhibits a more complex structure, which finds expression in three additional minima,
namely at $\sim-29$, $-7$, and 3~km\,s$^{-1}$. The radial velocities are measured relative to the
center of mass of the system. Thermal broadening is included by applying a constant temperature of
5000~K, which is approximately the mean temperature inside the \ion{H}{ii} region. Microturbulence
is not included in the calculation of the line profiles, so that all additional effects that
influence the line profile are of hydrodynamic origin.

\begin{figure}
 \resizebox{\hsize}{!}{\includegraphics[trim=0 0 0 0.5cm,clip=true]{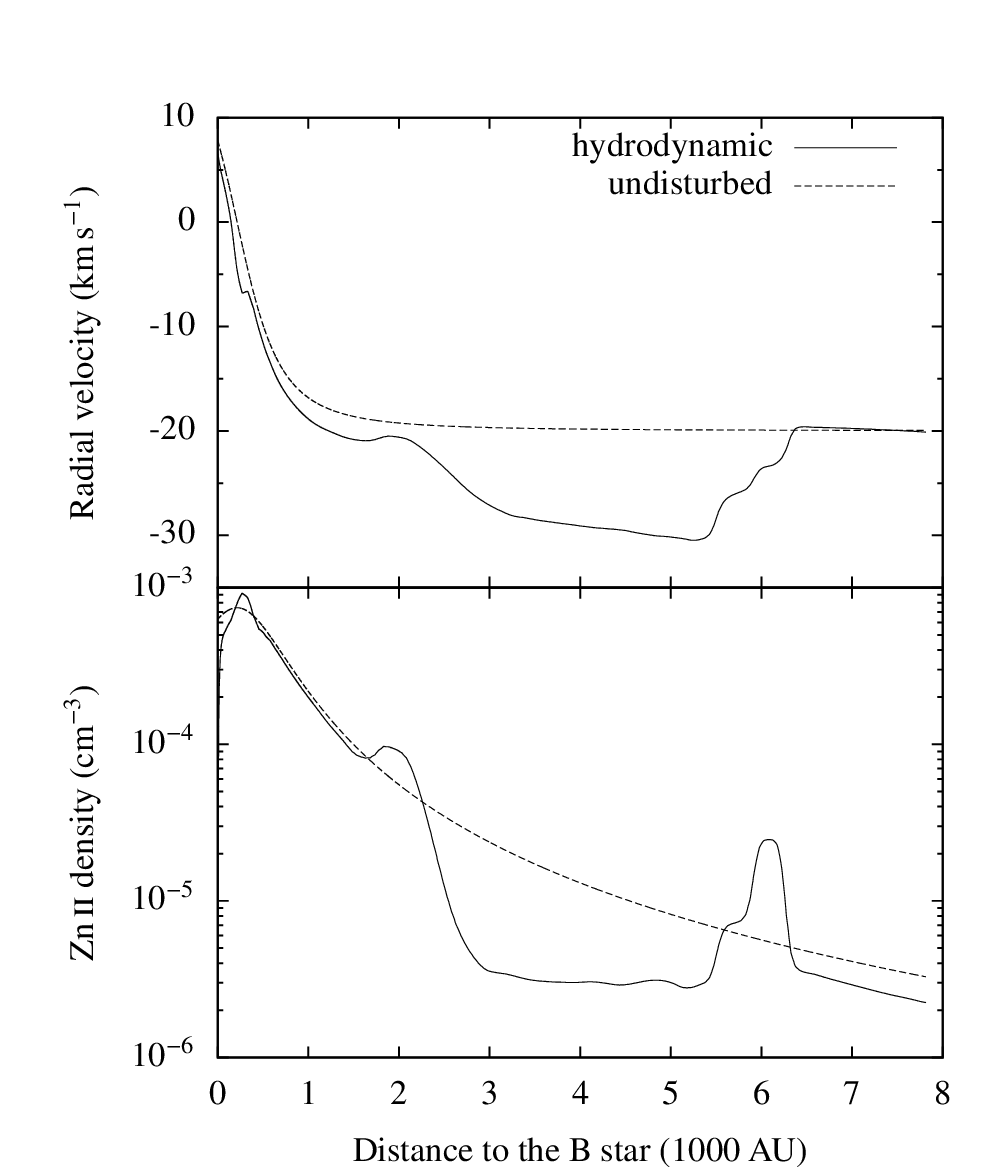}}
 \caption{Radial velocity relative to the center of mass and \ion{Zn}{ii} number density
          along the line of sight to the B star, based on a hydrodynamic simulation and on an
          undisturbed, spherically expanding wind. These data correspond to the absorption line
          profiles shown in Fig.~\ref{abspro} ($\dot{M}=10^{-6}\ M_\odot\,\mathrm{yr}^{-1}$,
          $\varv_{\infty}=20\ \mathrm{km\,s}^{-1}$).}
 \label{losqua}
\end{figure}

The distinct components seen in the absorption line presented in Fig.~\ref{abspro} are related to
features of the density and velocity distribution along the line of sight to the B star.
Figure~\ref{losqua} shows the number density of Zn$^{+}$ and the radial velocity as a function of
distance to the B star. One important result is that the velocity is no longer a monotonic function
of distance when hydrodynamic effects are included, but there is a broad minimum around
$\sim4000$~AU.

In order to test whether the simulations have converged, we calculated an additional line
profile based on a simulation with the same parameters, but with a resolution that is enhanced by a
factor of two (see Fig.~\ref{abspro}). The result does not show any significant changes. For this
high-resolution simulation we used a computational domain with a reduced size of
$a_x\times a_y\times a_z\sim6217\times6217\times3419$~AU, because it would have been
computationally too expensive to calculate a high-resolution model of the same size as the
low-resolution models. This reduced computational domain does not cover the whole high-velocity
region that produces the absorption component at $\sim-29\ \mathrm{km\,s}^{-1}$. That
means that the broad minimum in the radial velocity (Fig.~\ref{losqua}) is not fully covered, which
results in a weaker absorption component.

\begin{figure}
 \resizebox{\hsize}{!}{\includegraphics[trim=0 0 0 0.75cm,clip=true]{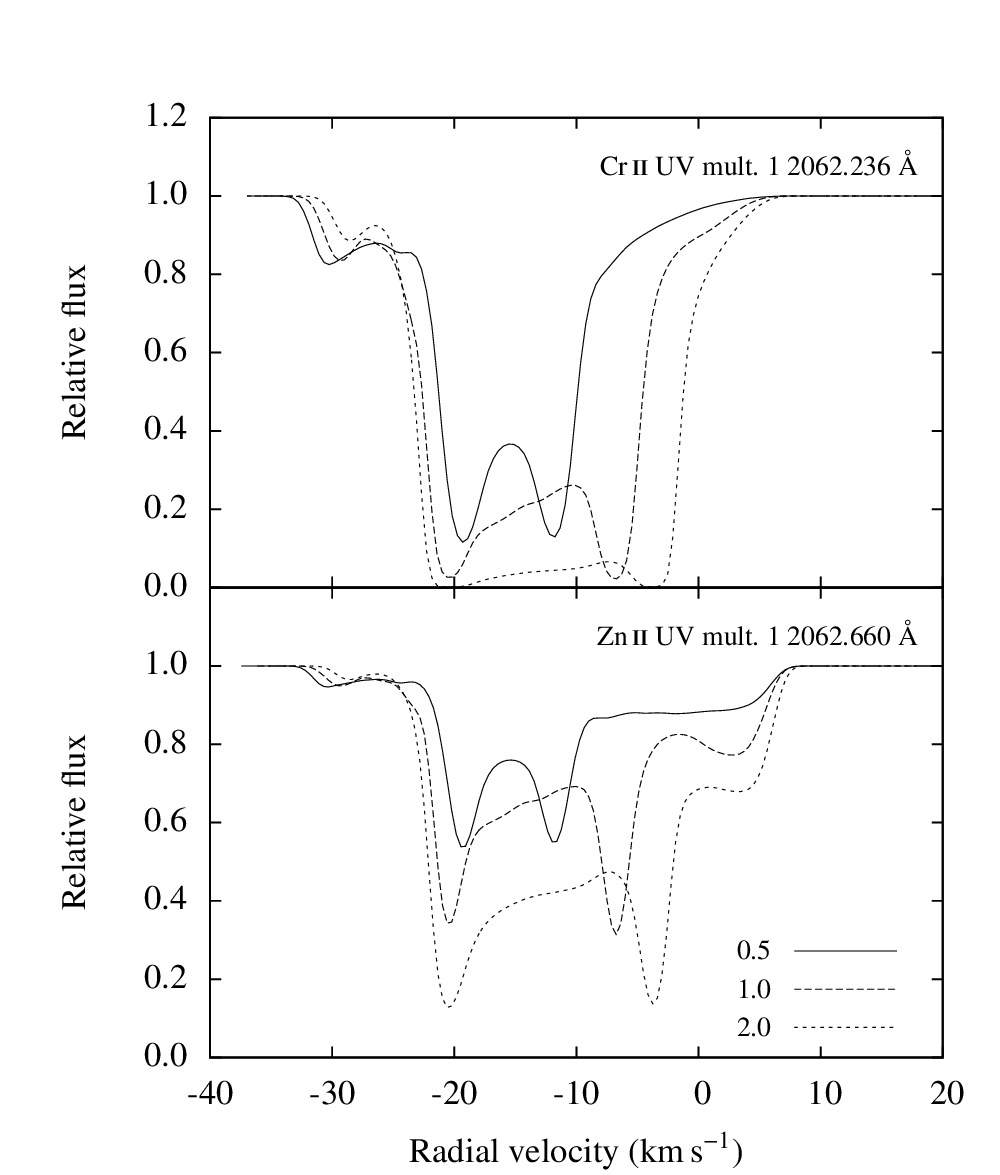}}
 \caption{Absorption profiles of the \ion{Zn}{ii} UV mult.~1 2062.660~{\AA} and \ion{Cr}{ii} UV
          mult.~1 2062.236~{\AA} lines for different mass-loss rates ($\varv_\infty=20\
          \mathrm{km\,s}^{-1}$). The mass-loss rates are given at the bottom right of the figure in
          units of $10^{-6}\ M_\odot\,\mathrm{yr}^{-1}$.}
 \label{mlrdep}
\end{figure}

Figure~\ref{mlrdep} shows the dependence of the \ion{Zn}{ii} UV mult.~1 2062.660~{\AA} and
\ion{Cr}{ii} UV mult.~1 2062.236~{\AA} lines on the mass-loss rate. While the position of the
component at $\sim-20\ \mathrm{km\,s}^{-1}$ remains nearly unchanged, the position of the other
strong component, which is associated to the increased density at the boundary of the \ion{H}{ii}
region, shifts to lower velocities as the mass-loss rate decreases and the \ion{H}{ii} region
becomes more extended. There is also a component at $\sim-29\ \mathrm{km\,s}^{-1}$, which is very
weak in the \ion{Zn}{ii} line. The profiles of the \ion{Cr}{ii} line show that this component is
considerably stronger at lower mass-loss rates. This is because the wake of the \ion{H}{ii} region
is more extended if the overall density is lower, so that the minimum of the radial velocity, shown
in Fig.~\ref{losqua} for $\dot{M}=10^{-6}\ M_\odot\,\mathrm{yr}^{-1}$, becomes broader and a higher
number of particles along the line of sight move with velocities larger than the terminal wind
velocity $\varv_{\infty}$. That is, the larger extent of the volume containing gas at high
velocities, which produces the absorption component at $-29\ \mathrm{km\,s}^{-1}$, overcompensates
the effect of the decreased density.

\begin{figure}
 \resizebox{\hsize}{!}{\includegraphics[scale=0.8]{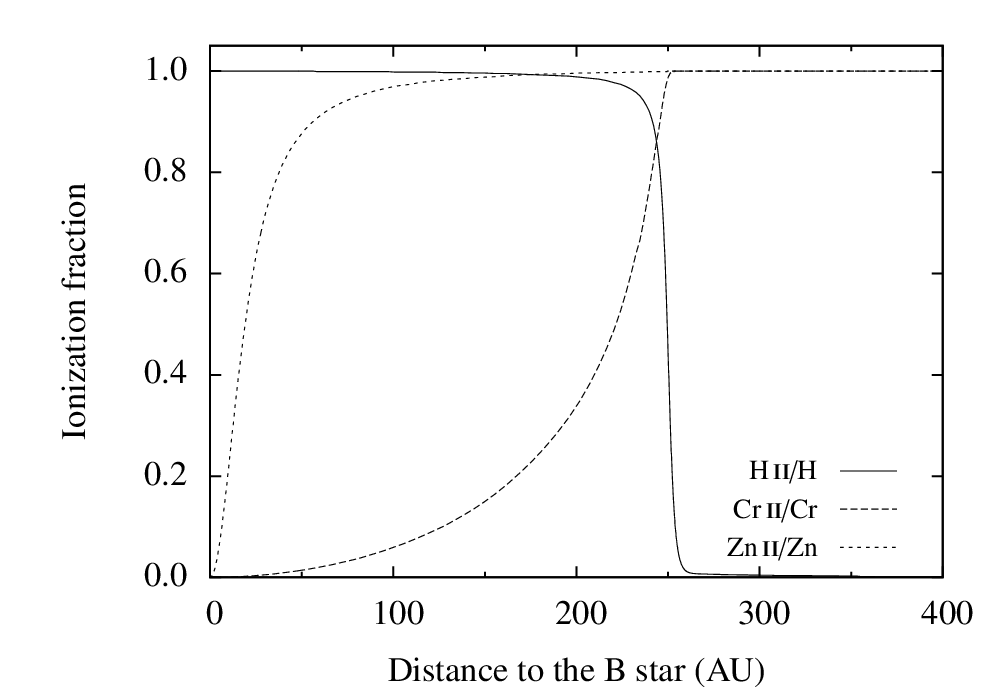}}
 \caption{Ionization fractions of different ions as a function of distance to the B star along the
          line of sight. The data are derived from a simulation using $\dot{M}=10^{-6}\
          M_\odot\,\mathrm{yr}^{-1}$ and $\varv_\infty=20\ \mathrm{km\,s}^{-1}$.}
 \label{crznio}
\end{figure}

In the \ion{Cr}{ii} line there is practically no absorption at positive velocities, because most
chromium is doubly ionized throughout the \ion{H}{ii} region. In contrast, Zn$^{++}$ is only present
close to the B star, which is partly due to the different ionization potentials of Zn$^{+}$
(144\,893~cm$^{-1}$) and Cr$^{+}$ (132\,966~cm$^{-1}$). The relative abundances of H$^{+}$,
Cr$^{+}$, and Zn$^{+}$ are shown in Fig.~\ref{crznio} as a function of distance to the B star along
the line of sight.

\subsubsection{Calibration of the observed wavelength scale} \label{calwav}
The positions of the observed absorption components in the GHRS spectra are slightly different in
every line \citep{Baade_Reimers_2007}, which is due to the relatively large error of the wavelength
calibration. The data were recorded using the Large Science Aperture of GHRS, which results in a
maximum error of $\sim4.5\ \mathrm{km\,s}^{-1}$ \citep{Heap_etal_1995}. The wavelength calibration
of the spectra observed with UVES is much more accurate. It amounts to $\ga120\ \mathrm{m\,s}^{-1}$
\citep{DOdorico_etal_2000}. The panel at the top right of Fig.~\ref{obsabs} shows the \ion{Ca}{ii}
mult.~1 (H and K) lines. There are three components in both lines, at $-21.0$, $-13.6$, and $-5.9\
\mathrm{km\,s}^{-1}$, respectively.

Now, supposing that the component at $-21\ \mathrm{km\,s}^{-1}$ in the \ion{Ca}{ii} lines is
associated to the terminal wind velocity as suggested by the synthetic line profiles presented in
the previous section, the positions of the GHRS lines can be calibrated by applying shifts that
place the most blue-shifted strong component at $-21\ \mathrm{km\,s}^{-1}$, which is acceptable as
long as the shifts are smaller than the maximum error of the GHRS wavelength calibration
(4.5~km\,s$^{-1}$). Table~\ref{calshi} lists the shifts corresponding to the five GHRS lines
presented in Fig.~\ref{obsabs}. The \ion{Cr}{ii} line is saturated, so that the positions of the
strong absorption components are rather uncertain. Therefore, the shift determined for the
\ion{Zn}{ii} line is used, which lies very close to the \ion{Cr}{ii} line. For the \ion{Fe}{ii} line
the central component is used to define the shift, because the observed component near $-21\
\mathrm{km\,s}^{-1}$ is weak and not clearly pronounced.

\begin{table}
 \caption{Velocity shifts applied to the GHRS spectra (see Fig.~\ref{obsabs}).}
 \label{calshi}
 \centering
 \begin{tabular}{l l}
  \hline\hline
  Transition & Shift (km\,s$^{-1}$) \\
  \hline
  \ion{Cu}{ii} UV mult.~3 1358.773~{\AA} & $-3.2$ \\
  \ion{Ni}{ii} 1393.324~{\AA}            & $-4.2$ \\
  \ion{Zn}{ii} UV mult.~1 2062.660~{\AA} & $-3.0$ \\
  \ion{Cr}{ii} UV mult.~1 2062.236~{\AA} & $-3.0$ \\
  \ion{Fe}{ii} UV mult.~1 2622.452~{\AA} & $-1.6$ \\
  \hline
 \end{tabular}
\end{table}

Observed absorption lines of \ion{Ti}{ii} also show a strong absorption component at $\sim-21\
\mathrm{km\,s}^{-1}$, which may serve as an additional justification for identifying the position of
this component with the terminal wind velocity. The two panels at the bottom right of
Figure~\ref{obsabs} show absorption line profiles of the \ion{Ti}{ii} multiplets 1 and 2.

\subsubsection{Observed line profiles} \label{obspro}
\begin{figure*}
 \centering
 \includegraphics[scale=0.1,width=17cm]{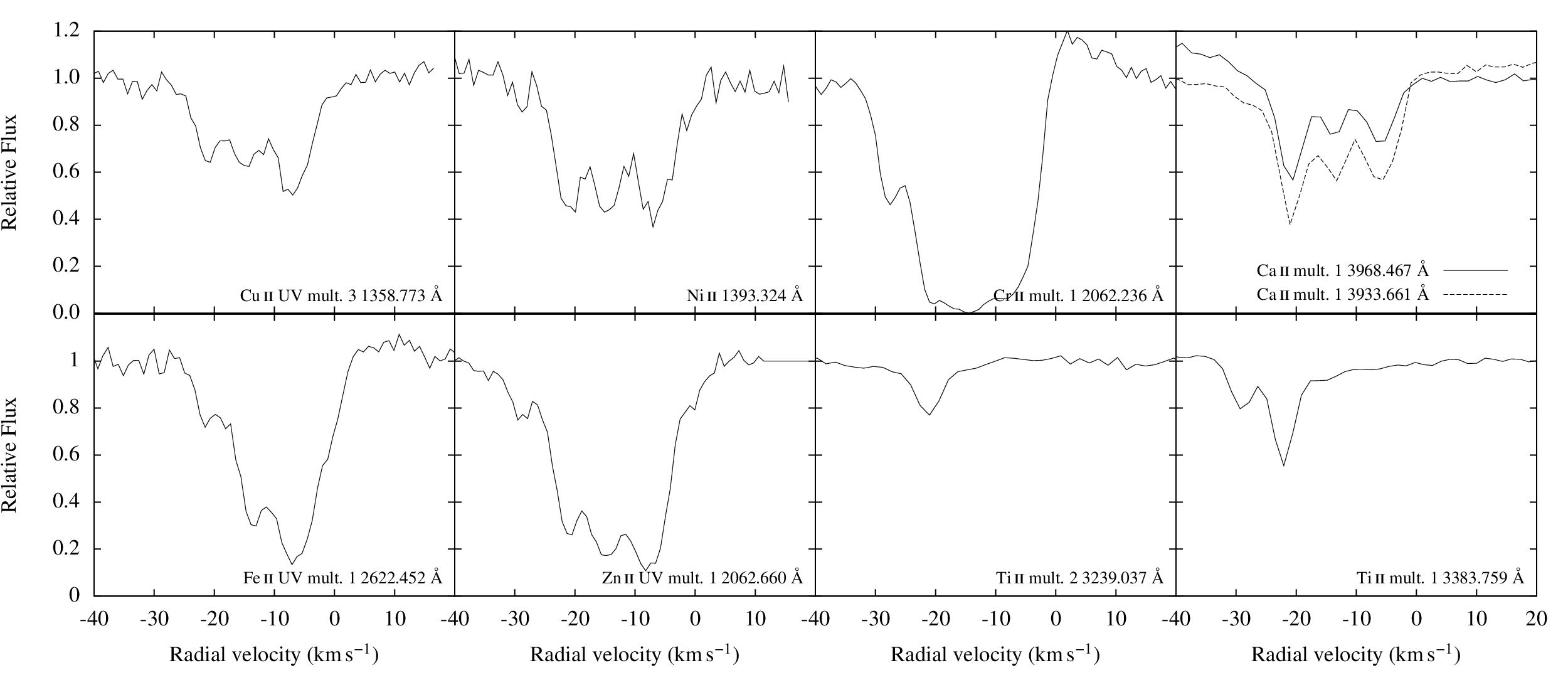}
 \caption{Absorption profiles observed in the spectrum of $\alpha$~Sco~B with HST/GHRS
          (\ion{Cu}{ii}, \ion{Fe}{ii}, \ion{Ni}{ii}, \ion{Zn}{ii}, and \ion{Cr}{ii} lines) and
          VLT/UVES (\ion{Ti}{ii} and \ion{Ca}{ii} lines).}
 \label{obsabs}
\end{figure*}

Figure~\ref{obsabs} shows a selection of absorption lines observed with GHRS/HST and UVES/VLT in the
spectrum of $\alpha$~Sco~B. The shapes of the line profiles produced by different ions or different
energy levels of the same ion (\ion{Ti}{ii}) show a large variety. The observed absorption features
depend on the excitation of fine-structure levels of the ions and the ionization of the
corresponding element along the line of sight. Most of the lines exhibit three strong absorption
components at the approximate positions $-21$, $-14$, and $-6\ \mathrm{km\,s}^{-1}$, respectively.
Some of the lines produced by the 0~eV levels have an additional component at $-29\
\mathrm{km\,s}^{-1}$ (\ion{Ti}{ii} 3383.759~\AA, \ion{Cr}{ii} 2062.236~\AA, \ion{Zn}{ii}
2062.660~\AA, and \ion{Ni}{ii} 1393.324~\AA), which can be explained by the broad minimum of the
radial velocity in the wake of the \ion{H}{ii} region far from the B star (see Figs.~\ref{mlrdep}
and \ref{losqua}).

For the comparison presented in the following, a systemic velocity of $\varv_{\mathrm{sys}}=-1.3\
\mathrm{km\,s}^{-1}$ is added to the wavelength scale of the simulated profiles, which results from
a comparison of the projected orbital velocities resulting from the adopted orbital configuration
($\delta=23$\degr, $e=0$) with the observed radial velocities \citep[see][]{Reimers_etal_2008}.

\begin{figure}
 \resizebox{\hsize}{!}{\includegraphics{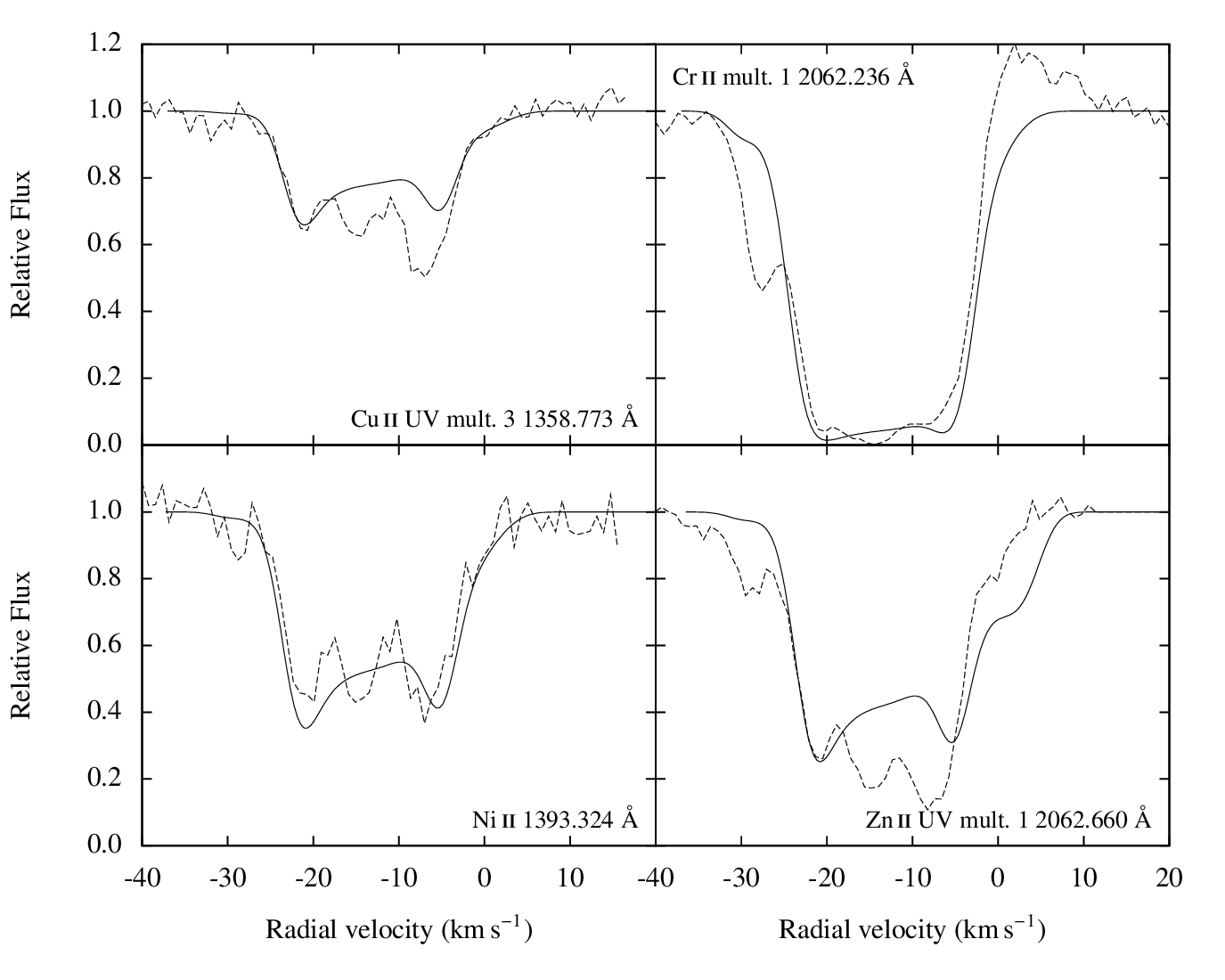}}
 \caption{Comparison of theoretical line profiles (solid curves) corresponding to a mass-loss rate
          of $\dot{M}=2\times10^{-6}\ M_\odot\,\mathrm{yr}^{-1}$ and a wind velocity of
          $\varv_\infty=20\ \mathrm{km\,s}^{-1}$ with observations (dashed curves), including a
          velocity calibration provided by the UVES spectra (see Sect.~\ref{calwav}) and a systemic
          velocity of $-1.3\ \mathrm{km\,s}^{-1}$. The relative flux is given as a function of the
          observed radial velocity.}
 \label{cnzcom}
\end{figure}

\paragraph{\ion{Cr}{ii}, \ion{Cu}{ii}, \ion{Ni}{ii}, and \ion{Zn}{ii}}
Figure~\ref{cnzcom} shows a comparison of the observed \ion{Cr}{ii}, \ion{Cu}{ii}, \ion{Ni}{ii}, and
\ion{Zn}{ii} lines presented in Fig.~\ref{obsabs} to theoretical profiles corresponding to a
mass-loss rate of $\dot{M}=2\times10^{-6}\ M_\odot\,\mathrm{yr}^{-1}$ and a wind velocity of
$\varv_\infty=20\ \mathrm{km\,s}^{-1}$. The simulated line profiles are convolved with a gaussian
profile function with a FWHM of 3.5~km\,s$^{-1}$ to account for the limited resolution of the
GHRS/HST spectra.

Obviously, there are a number of differences between the simulated and the observed line profiles.
First of all, there are only two strong components in the simulated profiles, whereas there are
three in the observed profiles. Secondly, the observed component at $-6\ \mathrm{km\,s}^{-1}$ is
much stronger than predicted by the model calculations in the \ion{Zn}{ii} and \ion{Cu}{ii} lines.
However, with the wavelength calibration described in Sect.~\ref{calwav} the positions of the two
strong components in the simulated profiles approximately match the positions of the outer two of
the observed strong components. It turns out that, within the error of the wavelength calibration,
the observed GHRS profiles are consistent with the assumption that the position of the most
blue-shifted of the strong absorption components is at $-21\ \mathrm{km\,s}^{-1}$, as suggested by
the absorption lines measured with UVES and by the theoretical lines, all of which include a strong
absorption component at the position corresponding to the terminal wind velocity.

The weaker absorption component at $\sim-29\ \mathrm{km\,s}^{-1}$, which is clearly visible in the
\ion{Cr}{ii} and the \ion{Zn}{ii} line, is very weak in the corresponding theoretical profiles.
The simulations predict a stronger absorption component at this position for lower mass-loss
rates, especially in the \ion{Cr}{ii} line (see Fig.~\ref{mlrdep}). This may be an indication of a
time-dependent mass-loss rate, which could produce distinct absorption components corresponding to
different mass-loss rates (see Sect.~\ref{astdep}).

The equivalent widths of the observed profiles and the simulated profiles shown in Fig.~\ref{cnzcom}
agree well. Moreover, the component at $\sim-21\ \mathrm{km\,s}^{-1}$ of the simulated profiles
gives a good fit of the observed components. The absorption in the components at $-6\
\mathrm{km\,s}^{-1}$ is stronger in the observed lines, which may be due to interstellar absorption
as all the lines result from transitions starting from the lowest energy levels of the ions. The
theoretical absorption lines resulting from simulations using a mass-loss rate of $\dot{M}=10^{-6}\
M_\odot\,\mathrm{yr}^{-1}$, as given by \citet{Reimers_etal_2008} based on a simplified model of the
circumstellar shell neglecting dynamic effects, are much weaker than the observed ones. That shows
the significant impact of the hydrodynamic effects, which will have to be considered in future
studies of mass loss.

Differences between the observed and simulated profiles are to be expected, because the simulations
do not account for a possible time dependence of the mass-loss rate. Indications of multiple shell
ejection have been observed in IR images in both $\alpha$~Sco and $\alpha$~Ori \citep[e.\,g.][see
also Sect.~\ref{introd}]{Danchi_etal_1994}. The additional third component in the observed lines
seen in Fig.~\ref{cnzcom} at $\sim-14\ \mathrm{km\,s}^{-1}$ could well be the result of such a
discrete ejection. An additional complication of a comparison of observations with theory is the
observed differential depletion \citep{Baade_Reimers_2007}. These effects may be the reason for a
more complicated structure of the absorption profiles.

\paragraph{\ion{Fe}{ii}}
The observed absorption profile of the \ion{Fe}{ii} UV mult.~1 2622.452~{\AA} line presented in
Fig.~\ref{obsabs} results from the fine structure level in the ground term at 977.053~cm$^{-1}$. The
population of this level cannot be reproduced by applying a constant scale factor to the total
Fe$^{+}$ density, as would be expected if the levels were populated according to their statistical
weights.

The model atom of \citet{Verner_etal_1999} that was used in the plasma simulations yields the
population of the level at 977.053~cm$^{-1}$, which is the highest level in the ground term. It
turns out that the relative population of the level is very low inside the cool neutral wind. The
population of the \ion{Fe}{ii} energy levels depends on the electron temperature, which determines
the rate of collisions. The temperature inside the \ion{H}{ii} region is $\sim5000$~K, but the
temperature in the \ion{H}{i} region is not exactly known.

\begin{figure}
 \resizebox{\hsize}{!}{\includegraphics{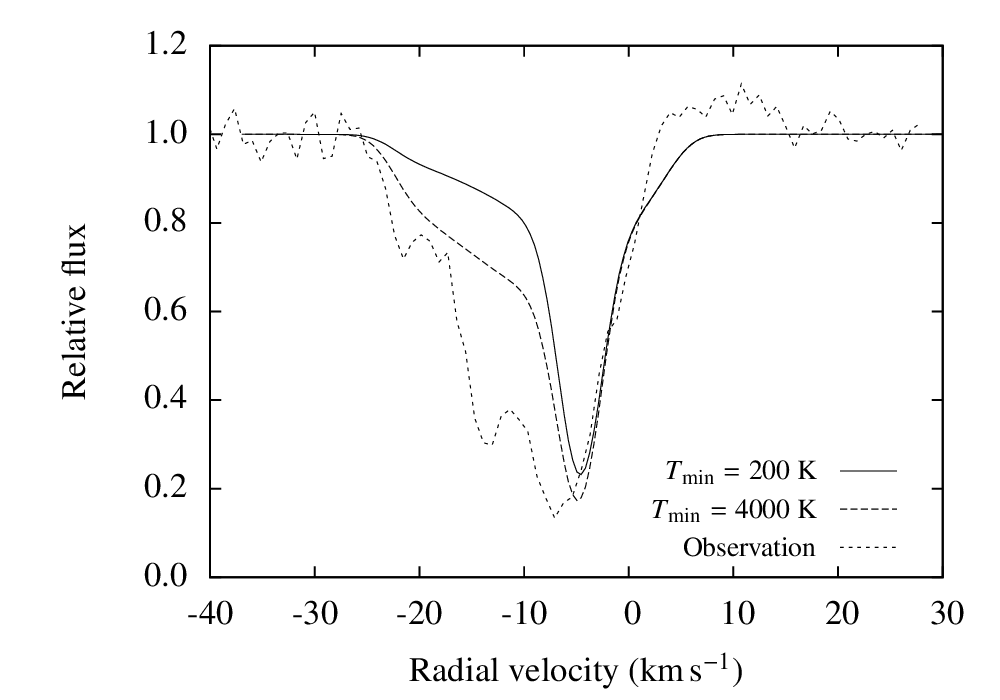}}
 \caption{Observed and theoretical absorption profiles of the \ion{Fe}{ii} UV mult.~1 2622.452~{\AA}
          line for $\dot{M}=2\times10^{-6}\ M_\odot\,\mathrm{yr}^{-1}$ and $\varv_\infty=20\
          \mathrm{km\,s}^{-1}$. For the theoretical profiles, two different lower limits of the
          electron temperature were used in the corresponding plasma simulations (see text). The
          relative flux is given as a function of the observed radial velocity.}
 \label{fe2abs}
\end{figure}

Figure~\ref{fe2abs} shows the absorption line profiles corresponding to two different lower limits
of the electron temperature along with the observed absorption profile, which is shifted by $-1.6\
\mathrm{km\,s}^{-1}$ according to Table~\ref{calshi} (cf.\ Sect.~\ref{calwav}). In these
simulations, the temperature was not allowed to fall below a given lower limit. Obviously, the
relative population of the 977.053~cm$^{-1}$ level steeply decreases after a peak at the boundary of
the \ion{H}{ii} region. The slope of the right flank of the strong absorption components in the
theoretical profiles does not exactly match the observed one, because there is a considerable amount
of reemission, which is not included in the calculation of the theoretical line profiles. The
decrease of the absorption towards lower velocities qualitatively matches the observation, although
the slope is much too steep in the theoretical profiles. The absence of the middle absorption
component in the theoretical profiles is not surprising since it is absent in all theoretical
profiles (cf.\ Sect.~\ref{synpro}).

The assumption that the neutral part of the wind has a temperature of 4000~K is probably not
realistic, and even in this case the predicted absorption is too weak in the outer parts of the
envelope. A value of $T\la1000$~K in the \ion{H}{i} region is more realistic. Apparently the
calculation of the level population with Cloudy does not give correct results, maybe because of an
underestimation of the rates of collisional or radiative processes that populate the level that is
responsible for the absorption, or because of advection effects, which are not included in the
simulations (see Sect.~\ref{discus}). In the plasma simulations, the gas is assumed to be at rest
with respect to the B star, which may introduce another uncertainty in the calculation of the
populations of fine-structure levels, as the velocity field may have considerable effects on the
continuum pumping rates. However, the strength of the absorption component at $\sim-6\
\mathrm{km\,s}^{-1}$ approximately matches the strength of the observed component, so that the
mass-loss rate of $2\times10^{-6}\ M_\odot\,\mathrm{yr}^{-1}$, which is consistent with the observed
profiles of the lines of other ions (cf.\ Fig.~\ref{cnzcom}), can be confirmed.

The results presented in this section suggest that the weakness of the observed absorption
components of this \ion{Fe}{ii} line at $-14$ and $-21\ \mathrm{km\,s}^{-1}$ is due to the
relatively high energy difference between the lowest level and the fine-structure level
corresponding to the transition. Apparently the mechanisms that excite \ion{Fe}{ii} to higher
fine-structure levels are not efficient in the outer parts of the circumstellar envelope, i.\,e.\ at
low densities, far from the B star and its \ion{H}{ii} region.

\paragraph{\ion{Ca}{ii} and \ion{Ti}{ii}}
The number densities of Ca$^{+}$ and Ti$^{+}$ along the line of sight cannot be reproduced in the
framework of the present model. In the case of Ti$^{+}$, the plasma simulations yield a density that
is much too high and density gradients that are too steep to reproduce the observed absorption
profiles. The ionization of Ti$^{+}$ is complicated because its ionization potential (13.576~eV) is
close to the Lyman edge. Simulations show that the ionization of Ti$^{+}$ sensitively depends on the
local temperature in the circumstellar shell. The absence of the absorption components at $-6$ and
$-14\ \mathrm{km\,s}^{-1}$ in the observed line profiles of \ion{Ti}{ii} shown in Fig.~\ref{obsabs}
suggests that most titanium is at least doubly ionized up to a large distance to the B star, i.\,e.\
even far outside the \ion{H}{ii} region. However, the absence of these components may also be due to
dust depletion. The component at $-30\ \mathrm{km\,s}^{-1}$ is observed only in the 0~eV transitions
of \ion{Ti}{ii}, which supports evidence from the simulations that the line is formed far away from
the system. In the case of Ca$^{+}$, the predicted number density is much too low. In the whole
circumstellar envelope Ca$^{++}$ appears to be the dominant stage of ionization of calcium, and
Ca$^{+}$ shows a very complex dependence on the varying conditions along the line of sight. Thus,
the \ion{Ca}{ii} lines are inappropriate for mass-loss diagnostics.

The simplified treatment of the corresponding model atoms may be the reason for the discrepancy
between theory and observation in the case of \ion{Ca}{ii} and \ion{Ti}{ii}. The Cloudy code treats
Ca$^{+}$ as a five-level atom, which includes the fine-structure levels of only the lowest three
terms (4s, 3d, and 4p) with the corresponding transitions (multiplets 1, 2, and 1F). Ti$^{+}$ is
treated in the framework of the scandium-like isoelectronic sequence as a five-level atom without
resolving any fine-structure components, including only the terms a$^{4}$F,
z$^{4}\mathrm{G}^{\mathrm{o}}$, z$^{4}\mathrm{F}^{\mathrm{o}}$, and the terms
$^{4}\mathrm{F}^{\mathrm{o}}$ and $^{4}\mathrm{D}^{\mathrm{o}}$ of the configuration
3d($^{2}$D)4s4p($^{3}\mathrm{P}^{\mathrm{o}}$). Thus, for Ti$^{+}$, populations of individual
fine-structure levels are not calculated. Moreover, only transitions to the ground term of
\ion{Ti}{ii} are included in the simulation. The depletion of titanium due to dust formation is
probably the most significant effect producing deviations from the observed data
\citep[cf.][]{Baade_Reimers_2007}.

\subsection{Comparison to optical line emission from the Antares nebula}
\begin{figure}
 \begin{center}
  \includegraphics{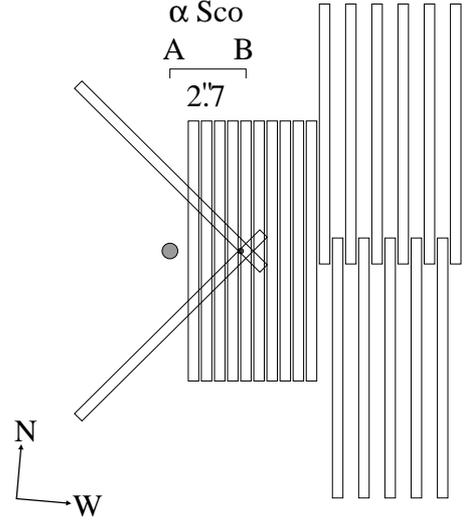}
 \end{center}
 \caption{Slit positions used for the observations of the Antares nebula. The rectangles indicate
          the position and size of the slit in the blue arm relative to the $\alpha$~Sco system as
          it is seen in the sky. This is Fig.~1 from \citet{Reimers_etal_2008}.}
 \label{slipos}
\end{figure}

The Antares emission nebula, which is associated to the \ion{H}{ii} region around $\alpha$~Sco~B,
is seen in the optical in H$\alpha$ and various other emission lines, last described in detail
by \citet{Reimers_etal_2008}. The simulations presented above will be compared to observations of
the H$\alpha$ emission in Sect.~\ref{haemis}. Moreover, forbidden transitions of \ion{Fe}{ii}, which
produce the most prominent emission lines of the Antares nebula, will be used to compare
observations to our model in Sect.~\ref{feemis}. The observations were performed with UVES/VLT using
a long slit perpendicular to the line connecting the two stars, and yielded a scan of the emission
nebula with a step size of $\sim0\farcs5$, starting 0\farcs9 west of the supergiant (see
Fig.~\ref{slipos}).

\subsubsection{H$\alpha$ emission} \label{haemis}
The plasma simulations show that the strongest H$\alpha$ emission in the Antares nebula emerges near
the B star, where most hydrogen is ionized. Outside the \ion{H}{ii} region the emission is much
weaker and follows the total density (Fig.~\ref{denvel}). In an \ion{H}{ii} region, H$\alpha$ is
naturally produced by recombination of electrons and protons. Another important mechanism is pumping
by Lyman line photons from the source of radiation, so that the presence of Lyman absorption or
emission lines in the spectrum of the source of radiation affects the production of H$\alpha$
photons. The B-star atmospheres presented in the TLUSTY grid \citep{Lanz_Hubeny_2007} show strong
absorption in the Lyman lines. As the Antares \ion{H}{ii} region is optically thin in H$\alpha$, the
observed emission is solely determined by the population of the upper levels and the oscillator
strengths. The sensitivity of the H$\alpha$ emission produced in the \ion{H}{ii} region to changes
of the input flux in the range of the Lyman lines requires a high-resolution input spectrum in the
Cloudy simulations. For the calculations presented in this section we used a resolution of
$\Delta\nu /\nu=5\times 10^{-4}$.

\begin{figure}
 \resizebox{\hsize}{!}{\includegraphics[trim=0 0 0 0.5cm,clip=true]{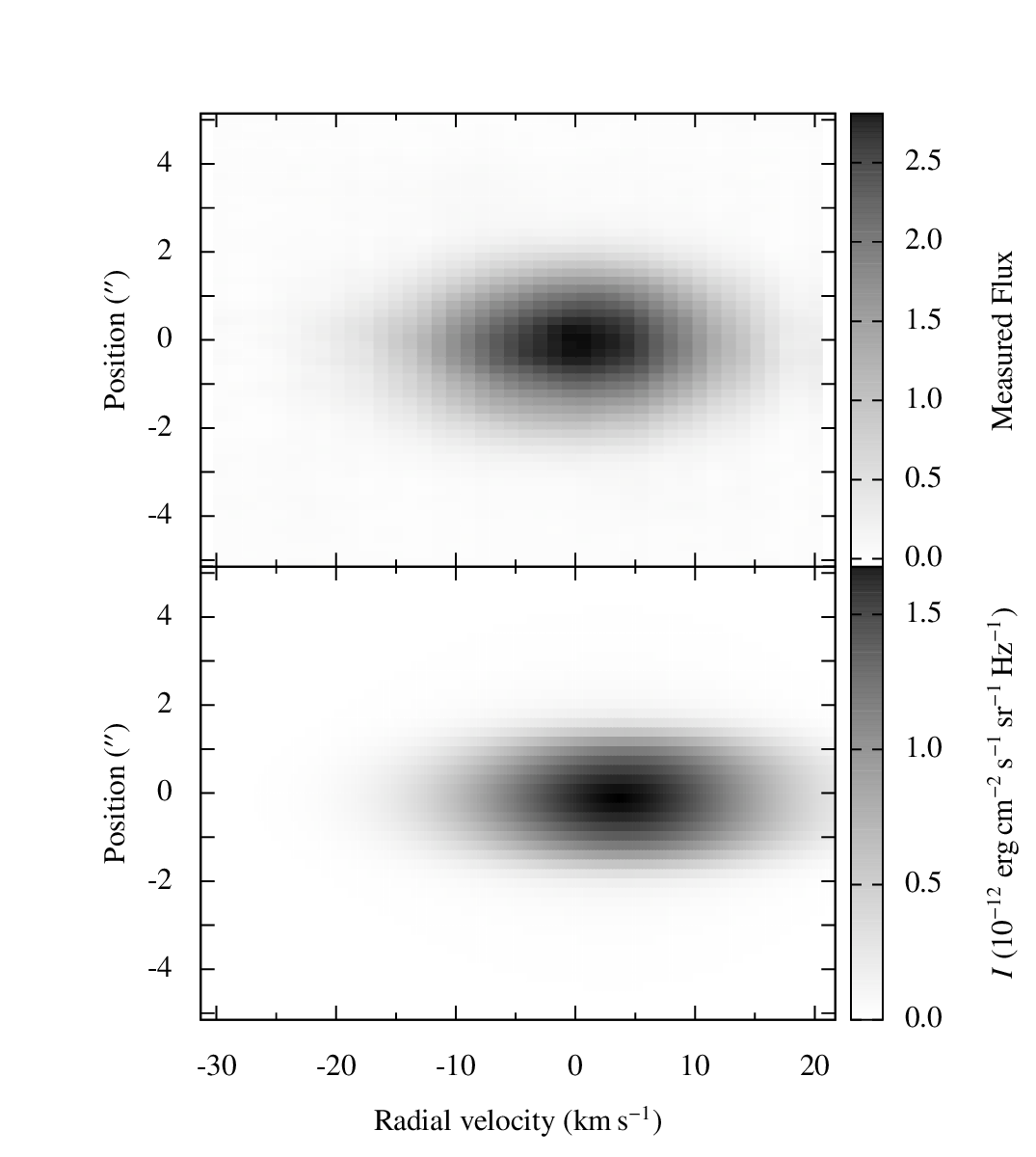}}
 \caption{H$\alpha$ emission at 3\farcs4 from the supergiant. The top panel shows the observed flux
          in arbitrary units, the bottom panel the intensity derived from a simulation with
          $\dot{M}=2\times10^{-6}\ M_\odot\,\mathrm{yr}^{-1}$ and $\varv_{\infty}=20\
          \mathrm{km\,s}^{-1}$. The ordinate indicates the position along the slit. As can be seen,
          cf.\ also Fig.~\ref{hadist}, the theorical distribution is slightly too narrow, which
          means that the adopted mass-loss rate is here slightly too large.}
 \label{hatwod}
\end{figure}

The spatial extent of the \ion{H}{ii} region is determined by the mass-loss rate of the supergiant
and the Lyman-continuum flux of the B star, and can be measured by observing the spatial extent of
the H$\alpha$ emission from the Antares nebula \citep[cf.][]{Reimers_etal_2008}. Figure~\ref{hatwod}
shows the H$\alpha$ intensity distribution at 3\farcs4 from the supergiant, measured on the
projected line connecting the two stars (see Fig.~\ref{slipos}). The shape and extent of the
observed and theoretical H$\alpha$ intensity distributions agree, but the center of the theoretical
distribution is shifted to higher velocities with respect to the observed distribution. This is
probably due to the uncertainties in the geometry of the system and can be explained by an
overestimation of the position angle of the line connecting the two stars relative to the plane of
the sky ($e=0$ is only a rough estimate, and due to the long period it will be difficult to improve
on this).

The intensity in Fig.~\ref{hatwod} is given as a function of velocity. Assuming pure emission, the
emergent intensity is given by the integral
\begin{equation}
 I_\varv(\varv)=\int_{s_\mathrm{min}}^{s_\mathrm{max}}j(s)\Phi[\varv_{r}(s)-\varv]\mathrm{d}s,
\end{equation}
where $s$ ranges from $s_\mathrm{min}=-0.3a$ to $s_\mathrm{max}=+0.3a$ along
the line of sight, and the zero point lies in the plane defined by the line connecting the two stars
and the direction perpendicular to the orbit, so that the \ion{H}{ii} region is fully included.
$a$ is the edge length of the domain in $x$ and $y$
direction. $s_\mathrm{min}$ and $s_\mathrm{max}$ were chosen such that the
integration includes a maximum of the available simulated data but does not exceed the boundaries of
the computational domain at the outermost slit position. As most of the H$\alpha$
emission comes from the \ion{H}{ii} region, this range ensures that the integration covers the whole
emitting region. $j$ is the emission coefficient for isotropic emission,
\begin{equation}
 j=\frac{h\nu_\mathrm{lu}n_\mathrm{u}A_\mathrm{lu}}{4\pi}\int_0^\infty\phi(\nu)\mathrm{d}\nu,
\end{equation}
resulting from the frequency $\nu_\mathrm{lu}$ of the transition, the number density $n_\mathrm{u}$
of ions in the upper state, the transition probability $A_\mathrm{lu}$, and the profile function
$\phi(\nu)$. $h$ is Planck's constant, $\varv_r(s)$ the radial velocity at $s$, and $\Phi$ a
gaussian profile that introduces thermal broadening,
\begin{equation}
 \Phi(\varv)=\frac{1}{\Delta\varv_\mathrm{D}\sqrt{\pi}}
                                     \exp{\left[-\frac{\varv^2}{(\Delta\varv_\mathrm{D})^2}\right]},
\end{equation}
where the Doppler width $\Delta\varv_\mathrm{D}$ is defined by pure thermal broadening with a
temperature of 5000~K. Microturbulent broadening is not included, so that the hydrodynamic effects
can be clearly identified. $I_\varv$ is the intensity per unit radial velocity, and the intensity
per unit frequency reads $I_\nu=\lambda_0I_\varv$, where $\lambda_0$ is the rest wavelength of the
transition.

For a comparison with the observations as presented in Fig.~\ref{hatwod} for H$\alpha$, the spectral
resolution of UVES and the seeing have to be considered. An analysis of the lines of the wavelength
calibration lamp yields an average FWHM of 3.5~km\,s$^{-1}$ in the vicinity of the [\ion{Fe}{ii}]
mult.~20F 4814.55~{\AA} line, which we present in the next section, in the range from 4789 to
4832~\AA. We adopt this value for the spectral resolution of the UVES spectra. The seeing during the
observations was $\sim0\farcs6$. Therefore, the theoretical intensity distributions are convolved
with a gaussian profile of $\mathrm{FWHM}=3.5\ \mathrm{km\,s}^{-1}$ along the frequency coordinate
and with another gaussian corresponding to the seeing in the direction of the spatial coordinate
along the slit. A systemic velocity of $-1.3\ \mathrm{km\,s}^{-1}$ is added to the theoretical
velocity scale (see Sect.~\ref{obspro}).

\begin{figure*}
 \centering
 \includegraphics[width=17cm]{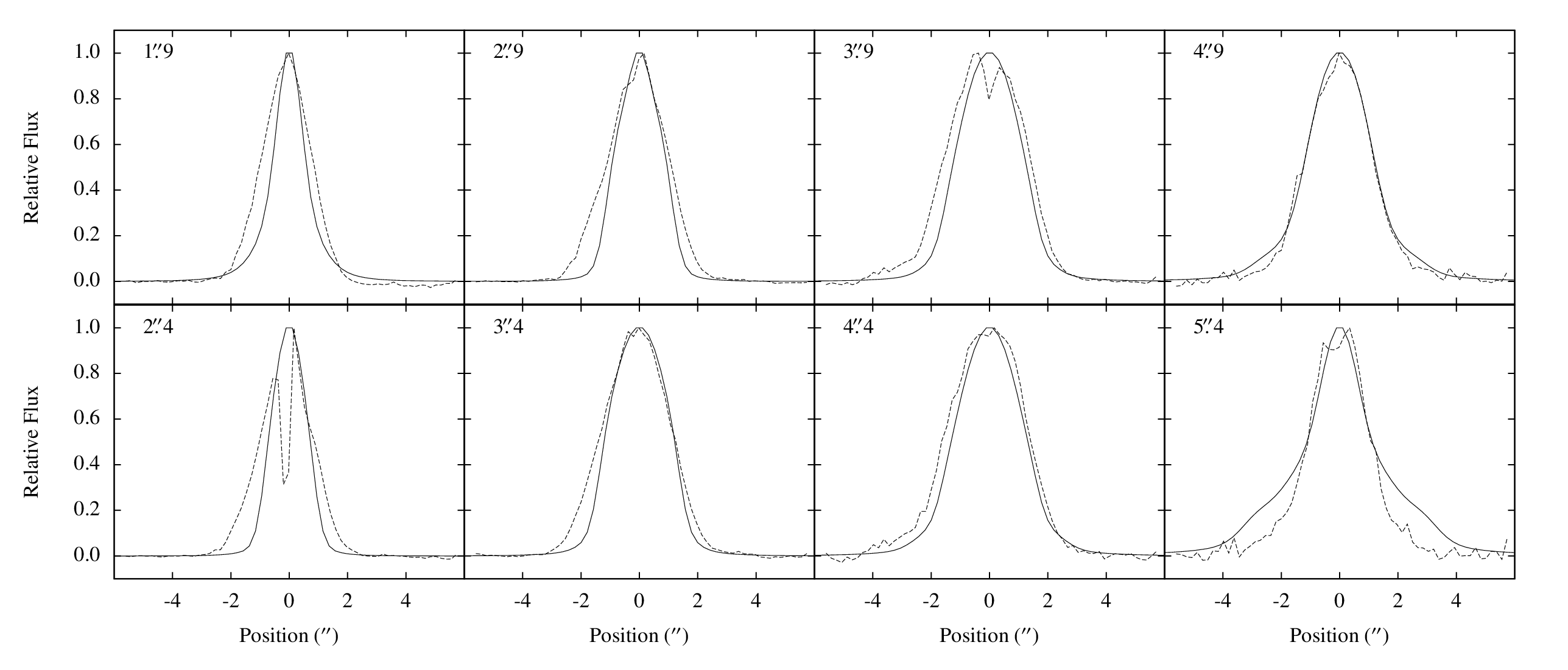}
 \caption{Frequency-integrated H$\alpha$ flux as a function of position along the slit for a
          simulation with $\dot{M}=2\times10^{-6}\ M_\odot\,\mathrm{yr}^{-1}$ and $\varv_\infty=20\ \mathrm{km\,s}^{-1}$ (solid lines), and the corresponding observations (dashed lines).}
 \label{hadist}
\end{figure*}

For a comparison of the spatial extent of the observed and theoretical emission, it is easier to
compare the frequency-integrated emission as a function of position along the slit. The integration
of the theoretical data reads
\begin{equation}
 I=\int_{s_\mathrm{min}}^{s_\mathrm{max}}j(s)\mathrm{d}s.
\end{equation}
The results are shown for different slit positions in Fig.~\ref{hadist} for a simulation with
$\dot{M}=2\times10^{-6}\ M_\odot\,\mathrm{yr}^{-1}$ and $\varv_\infty=20\ \mathrm{km\,s}^{-1}$,
along with the corresponding observational data. In this simulation, the slit skims only the
outermost part of the \ion{H}{ii} region at the slit positions 1\farcs9 and 5\farcs4. Therefore, the
theoretical data at these slit positions should be interpreted with care, because outside the
\ion{H}{ii} region the H$\alpha$ production is dominated by line pumping effects, which may not be
well reproduced with the simplified approach used by Cloudy for the radiative transfer (see
Sect.~\ref{plasim}). The central minimum in the observations corresponding to 2\farcs4 is probably
due to an artifact related to the data reduction. At the other slit positions, the simulated and
observed distributions agree well in shape and extent.

At 2\farcs4 and 2\farcs9, the theoretical H$\alpha$ profiles are narrower than the observed
profiles, indicating a lower mass-loss rate, while the observed data at the other slit positions
further away from the supergiant are better reproduced by the model. This may be due to the mass
loss rate being time-dependent.

\subsubsection{[\ion{Fe}{ii}] emission} \label{feemis}
The most prominent emission lines seen in the Antares nebula are the forbidden iron lines. The
spatial and spectral distribution of the line at 4814.55~{\AA} (multiplet F20) was presented in
detail by \citet{Reimers_etal_2008}. The data obtained with the Cloudy calculations, which include
the large Fe$^{+}$ model atom of \citet{Verner_etal_1999} (see Sect.~\ref{plasim}), can be used to
compare the results of the simulations to the observed data.

\begin{figure}
 \resizebox{\hsize}{!}{\includegraphics[trim=0 0 0 0.5cm,clip=true]{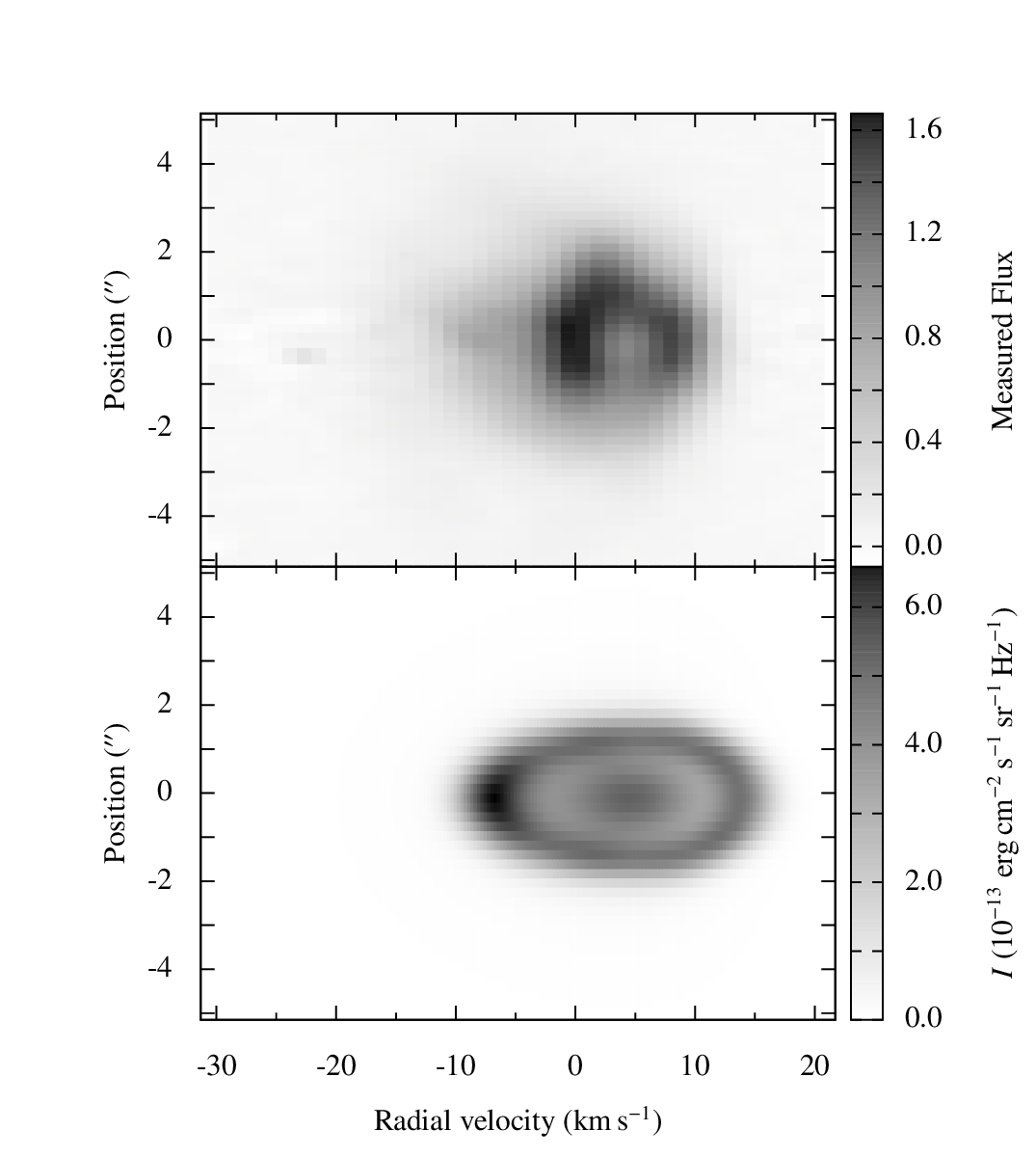}}
 \caption{[\ion{Fe}{ii}] mult.~20F 4814.55~{\AA} emission at 3\farcs4 from the supergiant. The top
          panel shows the observed flux in arbitrary units, the bottom panel the intensity derived
          from a simulation with $\dot{M}=2\times10^{-6}\ M_\odot\,\mathrm{yr}^{-1}$ and
          $\varv_\infty=20\ \mathrm{km\,s}^{-1}$. The ordinate indicates the position along the
          slit.}
 \label{fedist}
\end{figure}

As an example, Fig.~\ref{fedist} shows the distribution of the [\ion{Fe}{ii}] mult.~20F
4814.55~{\AA} intensity at 3\farcs4 both for the observed and the simulated data. Obviously, the
simulations do not yield a realistic picture of the [\ion{Fe}{ii}] emission for that slit position.
The simulations suggest a circular structure around a central maximum at $\sim5\
\mathrm{km\,s}^{-1}$, while the maximum flux in the observed data is concentrated in a more compact
structure, which is approximately circular but open to the bottom, between $\sim0$ and $\sim8\
\mathrm{km\,s}^{-1}$ as measured at the center of the slit. The overall extent of the observed
emission approximately agrees with the theoretical distribution and is consistent with the extent of
the H$\alpha$ emission in both the velocity and the spatial coordinate.

The discrepancies between the observed and theoretical [\ion{Fe}{ii}] emission could be due to an
inadequate treatment of the Fe$^{+}$ ion in the Cloudy code. The emission may also be influenced by
advection effects, which may be significant especially at the western boundary of the \ion{H}{ii}
region towards the open side of the wake, where the density is small. An alternative explanation
would be that due to dust depletion in the high density boundaries of the structure, a large
fraction of Fe is not available for \ion{Fe}{ii} emission. This effect seems to increase with
distance to the B star.

\begin{figure}
 \resizebox{\hsize}{!}{\includegraphics[trim=0 0 0 0.5cm]{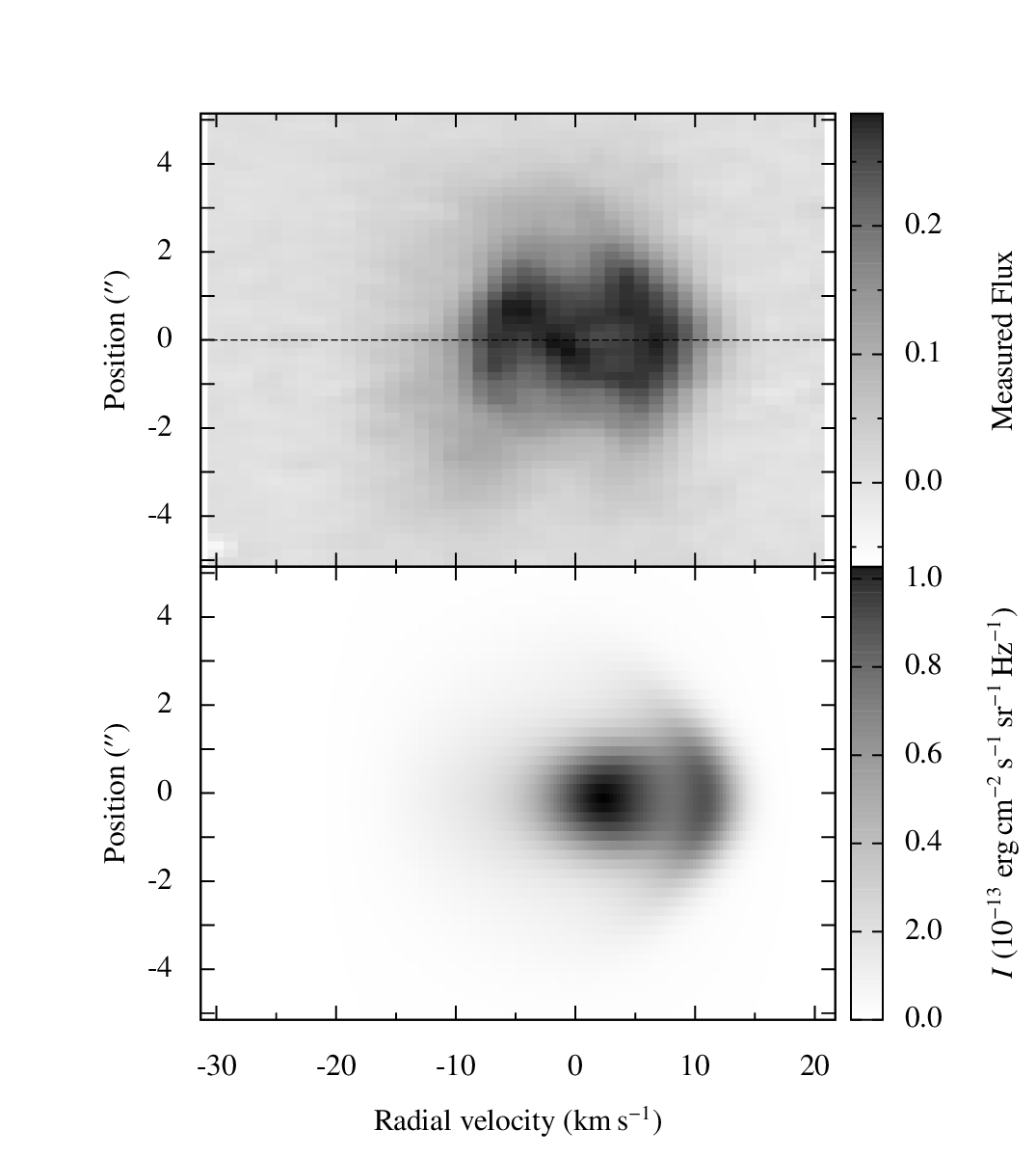}}
 \caption{[\ion{Fe}{ii}] mult.~20F 4814.55~{\AA} emission at 5\farcs4 from the supergiant. The top
          panel shows the observed flux in arbitrary units, the bottom panel the intensity derived
          from a simulation with $\dot{M}=2\times10^{-6}\ M_\odot\,\mathrm{yr}^{-1}$ and
          $\varv_{\infty}=20\ \mathrm{km\,s}$. The ordinate indicates the position along the slit.
          Obviously, there is more emission from the upper (northern) half in the observed flux
          distribution.}
 \label{feasym}
\end{figure}

A common feature of the observed forbidden and allowed Fe$^{+}$ emission is an asymmetry with
respect to the center of the slit, which is clearly visible e.\,g.\ at 5\farcs4 from the supergiant
(Fig.~\ref{feasym}). Apparently, there is more emission from the northern than from the southern
half (see also Fig.~\ref{fedist}). This is another indication of density structures that are not
included in the model, which is symmetric with respect to the plane of the orbit. A deviation of the
inclination from 90{\degr} would also cause such an asymmetry. However, the H$\alpha$ flux
distributions do not show any significant asymmetries.

\section{Discussion} \label{discus}

\subsection{The mass-loss rate}
The hydrodynamic simulations presented in Sect.~\ref{hydeve} show that the hot \ion{H}{ii} region
that is moving with the B star through the circumstellar envelope of $\alpha$~Sco produces strong
deviations from spherical symmetry in the density and velocity distributions. The theoretical
absorption line profiles derived from these simulations (Sect.~\ref{synpro}) exhibit a pronounced
multi-component structure comparable to the observed line profiles presented in Fig.~\ref{obsabs}.
The best match between theoretical and observed line profiles is achieved with a mass-loss rate of
$\dot{M}=2\times10^{-6}\ M_\odot\,\mathrm{yr}^{-1}$ and a wind velocity of $\varv_\infty=20\
\mathrm{km\,s}^{-1}$ (Sect.~\ref{obspro}). This value of the mass-loss rate is twice as high as the
rate derived by \citet{Reimers_etal_2008}, which was based on a spherically expanding model of the
circumstellar shell. This discrepancy is due to the decreased density in the wake of the \ion{H}{ii}
region (see Figs.~\ref{denvel} and~\ref{denzoo}).

The range of values of $\dot{M}$ and $\varv_\infty$ that is covered by the simulations is rather
small, owing to the limitations imposed by the large amount of computing time that the hydro code
requires. As the AMRCART code is not adapted to modern parallel computing clusters, the hydrodynamic
simulations were carried out on a single CPU core, while only the photoionization/radiative-transfer
calculations were executed in parallel. A finer grid in $\dot{M}$ and $\varv_\infty$ would improve
the accuracy of the resulting mass-loss rate, and it might be instructive to calculate models with
different semi-major axes and position angles to account for the uncertainties in the orbital
parameters (see next section). However, the theoretical data presented in this work and the
available observations agree fairly well, and the remaining systematic errors (see below)
are probably larger than the error that is due to the coarseness of the grid of models,
i.\,e.\ the relatively large step size in $\dot M$ and $\varv_\infty$. We
presume that the total error of the mass-loss rate including systematic uncertainties is
$\sim5\times10^{-7}\ M_\odot\,\mathrm{yr}^{-1}$.

An important systematic error is introduced by the observed differential dust depletion
\citep{Baade_Reimers_2007}. This effect depends on the considered element and on the local
conditions in the circumstellar shell. It results in the observed absorption profiles being weaker
than predicted for a given mass-loss rate. In contrast, interstellar absorption can make the
absorption components near 0~km\,s$^{-1}$ stronger than predicted by the model, which is probably
responsible for the discrepancies in the \ion{Zn}{ii} 2062.660~{\AA} and \ion{Cu}{ii} 1358.773~{\AA}
lines (Fig.~\ref{cnzcom}). According to \citet{Snow_etal_1987} and \citet{Cardelli_1984}, dust
depletion is of minor importance for zinc, which means that the mass-loss rate can best be
determined by use of the \ion{Zn}{ii} absorption lines.

In the present approach, the radiative transfer is treated in a simplified manner with the escape
probability formalism (cf.\ Sect.~\ref{plasim}). This is probably the major source of error in the
calculation of the [\ion{Fe}{ii}] line emission, and the exact solution of the scattering problem
would probably resolve some of the apparent discrepancies between the theoretical predictions and
the observations. Moreover, some of the circumstellar absorption lines in the spectrum of
$\alpha$~Sco~B, e.\,g.\ \ion{Cr}{ii} 2062.236~{\AA}, have P~Cyg-type profiles, which cannot be
reproduced with a pure absorption model. For a calculation of the reemission that is superposed in
these lines on the absorption profile, exact radiative transfer simulations have to be included in
the model.

\subsection{Asymmetries and time-dependent effects} \label{astdep}
The observations indicate an asymmetric, non-stationary character of the circumstellar shell of
$\alpha$~Sco. The observed absorption lines exhibit an additional component at $-14\
\mathrm{km\,s}^{-1}$, which cannot be reproduced with the assumption of a constant mass-loss rate.
Moreover, the integrated H$\alpha$ flux (Fig.~\ref{hadist}) also appears to indicate slightly
different mass-loss rates at different distances to the B star (Sect.~\ref{haemis}). The
observations of [\ion{Fe}{ii}] emission reveal that the density distribution is not symmetric with
respect to the plane of the orbit, which cannot be reproduced in the framework of the present model.

The uncertainties in the geometry of the $\alpha$~Sco system may partly account for the asymmetry
of the observed [\ion{Fe}{ii}] emission and the shift of the theoretical emission in H$\alpha$ and
[\ion{Fe}{ii}] 4814.55~\AA. The orbital parameters, such as the inclination, eccentricity, and
the orbital velocities, are not well known because of the large orbital period ($\sim2560$~yrs).
Especially the radial velocity of the supergiant is hard to measure, because the observed velocity
results from a superposition of pulsations and orbital motion \citep{Smith_etal_1989}.

The plasma calculations used for determining the temperature and ionization structure of the
circumstellar envelope are based on the assumption that the time scales of cooling and heating are
small compared to the dynamic time-scale. This is not exact at the western boundary of the
\ion{H}{ii} region, towards the low-density wake, where also advection effects may have a
significant impact on the ionization balance. A model including time-dependent plasma-effects may
considerably improve the understanding of the observed absorption and emission features and could be
the subject of future studies.

\section{Conclusions and Outlook}
The results of the combination of observations with hydrodynamic and plasma simulations presented in
this work show that the circumstellar envelope of the $\alpha$~Sco system is strongly influenced by
dynamic effects. A calculation of absorption line profiles based on the simulated density and
velocity distributions and ionization structure, and a comparison to GHRS/HST and UVES/VLT spectra
reveal that the mass-loss rate was underestimated by a factor of two in earlier studies that were
based on an undisturbed, spherically symmetric circumstellar shell. The resulting mass-loss rate of
$\dot{M}=2\times10^{-6}\ M_\odot\,\mathrm{yr}^{-1}$ is confirmed by an analysis of the H$\alpha$
emission from the Antares nebula, which is based on spatially resolved emission distributions
observed with UVES.

The theoretical absorption line profiles exhibit a multi-component structure as a natural result of
the influence of the hot \ion{H}{ii} region that is moving with the B star through the wind of the
primary. Three of the four observed components can be explained accordingly. However, the origin of
the additional component at $-14\ \mathrm{km\,s}^{-1}$ seen in most absorption lines
(Fig.~\ref{obsabs}) remains uncertain and is probably an indication of time-dependent mass-loss.

The structure of the [\ion{Fe}{ii}] line emission of the Antares nebula as observed with UVES cannot
be reproduced correctly. A more sophisticated model including exact radiative transfer calculations
and time-dependent simulations of the ionization balance allowing for advection effects would help
to understand the remaining discrepancies between theory and observations. In this context, it would
be desirable to parallelize the hydro code, which will make it possible to calculate a more
extensive grid of models. Future projects may deal with these improvements of the model
calculations.

A number of empirical mass-loss rates of $\zeta$~Aur systems were determined on the basis of
spherically symmetric density and velocity distributions \citep[see e.\,g.][]{Che_etal_1983,
Baade_etal_1996}. Some of these systems, e.\,g.\ $\zeta$~Aur and 31~Cyg, also contain \ion{H}{ii}
regions, which may lead to even more severe dynamic effects than in $\alpha$~Sco due to the much
smaller orbital periods. The current values of the mass-loss rates of these stars may have to be
revised on the basis of more realistic models including dynamic processes as presented in this work
for $\alpha$~Sco.

\begin{acknowledgements}
 K.~B.\ gratefully acknowledges support by a fellowship from the Landesgraduiertenf\"orderung
 Hamburg.
\end{acknowledgements}

\bibliographystyle{aa}
\bibliography{hydsco}

\appendix

\section{The temperature grid} \label{clodis}
For the calculation of the 3D temperature distribution, the 1D models calculated with Cloudy
are distributed homogeneously in 3D space, i.\,e.\ with a constant increment in $\theta$.
The B star is the starting point for all 1D models, which give the physical conditions along a
straight line in radial direction. The angle $\theta$ is measured from the upper pole,
i.\,e.\ from the positive $z$ direction in the AMRCART coordinate system, downwards, and
the angle $\phi$ is measured counterclockwise from the direction towards the supergiant.
The increment in $\phi$ in the plane of the orbit, i.\,e.\ at $\theta=\pi/2$,
equals the increment in $\theta$, which is determined by a given number $N$
according to $\Delta\theta=\pi/(N-1)$. Only odd numbers are used for $N$ so
that there are always two models describing the temperature distribution on the line connecting the
two stars, one away from the primary star, one towards it. In directions nearer to the poles,
$\Delta\phi$ is increased according to $\Delta\phi=\pi/(N\sin{\theta}-1)$,
where the expression $N\sin{\theta}$ is rounded to the nearest integer that is
$\geq2$. For $\theta=0$ and $\theta=\pi$ one direction
($\phi=0$) is calculated.

Thus, the total number of Cloudy simulations computed to cover the whole domain for a given
$N$ is
\begin{equation} \label{nclone}
 N_{\mathrm{Cl}}=2+\sum_{i=1}^{N-2}2(N_{i}-1),
\end{equation}
where $2(N_{i}-1)$ is the number of directions calculated for a given
$\theta_i=i\Delta\theta$, and
\begin{equation} \label{ncltwo}
 N_{i}=
  \begin{cases}
   N\sin{\theta_{i}}, & \sin{\theta_{i}}\geq2/N \\
   2,                 & \mathrm{else}
  \end{cases}.
\end{equation}

As the Cloudy simulations are independent of each other, they can be safely executed on
different CPUs. The simulations presented in this work are symmetric with respect to the orbital
plane so that it is sufficient to cover only the lower half of the domain, i.\,e.
\begin{equation} \label{nclsym}
 N_{\mathrm{Cl}}^{\mathrm{sym}}=1+\sum_{i=(N-1)/2}^{N-2}2(N_{i}-1).
\end{equation}
For most of the simulations presented in this work, we used $N=9$, i.\,e.\
$N_{\mathrm{Cl}}^{\mathrm{sym}}=45$. The only exception is the high-resolution simulation
(cf.\ Fig.~\ref{abspro}), where $N=13$ and
$N_{\mathrm{Cl}}^{\mathrm{sym}}=101$.

\end{document}